%Paper: cond-mat/9406119
%From: abdalla@surya11.cern.ch (Elcio Abdalla)
%Date: Thu, 30 Jun 94 09:32:35 +0200

%%%%%%%%%%%%%%%%%%%%%%%%%%%%%%%%%%%%%%%%%%%%%%%%%%%%%%%%%%%%%%%%%%%%%%%%%%%%%%

\magnification 1200

%\magnification = 1200
%

%
\font\eightrm=cmr8
\font\eighti=cmmi8
\font\eightsy=cmsy8
\font\eightbf=cmbx8
\font\eighttt=cmtt8
\font\eightit=cmti8
\font\eightsl=cmsl8
\font\sixrm=cmr6
\font\sixi=cmmi6
\font\sixsy=cmsy6
\font\sixbf=cmbx6
\catcode`@11
\newskip\ttglue
\font\grrm=cmbx10 scaled 1200

\def\eightpoint{\def\rm{\fam0\eightrm}
\textfont0=\eightrm \scriptfont0=\sixrm \scriptscriptfont0=\fiverm
\textfont1=\eighti \scriptfont1=\sixi \scriptscriptfont1=\fivei
\textfont2=\eightsy \scriptfont2=\sixsy \scriptscriptfont2=\fivesy
\textfont3=\tenex \scriptfont3=\tenex \scriptscriptfont3=\tenex
\textfont\itfam=\eightit \def\it{\fam\itfam\eightit}
\textfont\slfam=\eightsl \def\sl{\fam\slfam\eightsl}
\textfont\ttfam=\eighttt \def\tt{\fam\ttfam\eighttt}
\textfont\bffam=\eightbf
\scriptfont\bffam=\sixbf
\scriptscriptfont\bffam=\fivebf \def\bf{\fam\bffam\eightbf}
\tt \ttglue=.5em plus.25em minus.15em
\normalbaselineskip=6pt
\setbox\strutbox=\hbox{\vrule height7pt width0pt depth2pt}
\let\sc=\sixrm \let\big=\eightbig \normalbaselines\rm}
\newinsert\footins
\def\newfoot#1{\let\@sf\empty
  \ifhmode\edef\@sf{\spacefactor\the\spacefactor}\fi
  #1\@sf\vfootnote{#1}}
\def\vfootnote#1{\insert\footins\bgroup\eightpoint
  \interlinepenalty\interfootnotelinepenalty
  \splittopskip\ht\strutbox % top baseline for broken footnotes
  \splitmaxdepth\dp\strutbox \floatingpenalty\@MM
  \leftskip\z@skip \rightskip\z@skip
  \textindent{#1}\footstrut\futurelet\next\fo@t}
\def\fo@t{\ifcat\bgroup\noexpand\next \let\next\f@@t
  \else\let\next\f@t\fi \next}
\def\f@@t{\bgroup\aftergroup\@foot\let\next}
\def\f@t#1{#1\@foot}
\def\@foot{\strut\egroup}
\def\footstrut{\vbox to\splittopskip{}}
\skip\footins=\bigskipamount % space added when footnote is present
\count\footins=1000 % footnote magnification factor (1 to 1)
\dimen\footins=8in % maximum footnotes per page

\def\ref#1{$^{#1}$}
\def\flex{\raise 6pt\hbox{$\leftrightarrow $}\! \! \! \! \! \! }
\def\oversome#1{ \raise 8pt\hbox{$\scriptscriptstyle #1$}\! \! \! \! \! \! }
\def\tr{ \mathop{\rm tr}}

\newbox\bigstrutbox
\setbox\bigstrutbox=\hbox{\vrule height10pt depth5pt width0pt}
\def\bigstrut{\relax\ifmmode\copy\bigstrutbox\else\unhcopy\bigstrutbox\fi}
\def\refer[#1/#2]{ \item{#1} {{#2}} }
\def\rev<#1/#2/#3/#4>{{\it #1\/} {\bf#2}, {#3}({#4})}
\def\boxit#1{\vbox{\hrule\hbox{\vrule\kern3pt
\vbox{\kern3pt#1\kern3pt}\kern3pt\vrule}\hrule}}

\def\2figure#1#2#3#4{\vbox{ \hrule width#1truecm \hbox{\vrule height#2truecm
\hskip #1truecm
\vrule height#2truecm }\hrule width#1truecm \hbox{\vrule\vbox{\hsize #1truecm
\baselineskip=10pt
\noindent\strut#3}\vrule}\hrule width#1truecm
\hbox{\vrule\vbox{\hsize #1truecm
\baselineskip=10pt
\noindent\strut#4}\vrule}\hrule width#1truecm  }}
\def\3figure#1#2#3#4#5{\vbox{ \hrule width#1truecm \hbox{\vrule height#2truecm
\hskip #1truecm
\vrule height#2truecm }\hrule width#1truecm \hbox{\vrule\vbox{\hsize #1truecm
\baselineskip=10pt
\noindent\strut#3}\vrule}\hrule width#1truecm
 \hbox{\vrule\vbox{\hsize #1truecm
\baselineskip=10pt
\noindent\strut#4}\vrule}
\hrule width#1truecm \hbox{\vrule\vbox{\hsize #1truecm
\baselineskip=10pt
\noindent\strut#5}\vrule}\hrule width#1truecm  }}

\def\sqr#1#2{{\vcenter{\hrule height.#2pt
   \hbox{\vrule width.#2pt height#1pt \kern#1pt
    \vrule width.#2pt}
    \hrule height.#2pt}}}
\def\dal{\mathchoice{\sqr{6}{4}}{\sqr{5}{3}}{\sqr{5}3}{\sqr{4}3} \, }

% Here are my additional definitions:

\def\smin{\,\raise 0.06em \hbox{${\scriptstyle \in}$}\,}
\def\smsubset{\,\raise 0.06em \hbox{${\scriptstyle \subset}$}\,}

\def\Natural{\hbox{\hskip 1.5pt\hbox to 0pt{\hskip -2pt I\hss}N}}

\def\Rational{\hbox{\hbox to 0pt{\hskip 2.7pt \vrule height 6.5pt
                                  depth -0.2pt width 0.8pt \hss}Q}}
\def\Real{\hbox{\hskip 1.5pt\hbox to 0pt{\hskip -2pt I\hss}R}}
\def\Complex{\hbox{\hbox to 0pt{\hskip 2.7pt \vrule height 6.5pt
                                  depth -0.2pt width 0.8pt \hss}C}}
%%%%%%%%%%%%%%%%%
% definitions for the second book
%tcap1

%%%%%%%%tcap2

%%%%%%%%tcap3
\def \1ok{{1\over \kappa ^2} }

\def \3dslim {{\rm DS}\!\!\!\!\!\!\!\!\lim }
\def \4dslim {{\rm DS}\!\!\!\!\!\!\!\!\!\!\lim }
\def \tr {{\rm tr}\, }

\def \2kk{\left( \matrix {2k\cr k\cr }\right) }
\def \Rs4{{R^k\over 4^k} }

%%%%%%%%%tcap4
\def \1ok{{1\over \kappa ^2} }

%%%%%%%%%%tcap5

%%%%%%%%%%tcap6
%\def \DSLim {{\rm DS}\!\!\lim }
%%%%%%%%%%tcap7
%nada
%%%%%%%%%%tcap8
%nada
%%%%%%%%%appb

%\input macrosbk.tex
\nopagenumbers

\centerline{\grrm Field theory formulation of the Quantum Hall Effect}
\vskip 1.5cm
\centerline { E. Abdalla\newfoot {${}^*$}{Permanent address: Instituto de
F\'\i sica - USP, C.P. 20516, S. Paulo, Brazil.}, M.C.B. Abdalla\newfoot
{${}^\dagger $}{Permanent address: Instituto de F\'\i sica
Te\'orica - UNESP, R. Pamplona 145, 01405-000, S. Paulo, Brazil.} }
\vskip 1.5truecm

\centerline { CERN, Theory Division, CH-1211 Gen\`eve 23, Switzerland}
\vskip 2cm

\centerline {\bf Abstract}
\vskip .5cm
\noindent We consider the quantum Hall effect in terms of an effective field
theory formulation of the edge states, providing a natural common framework
for the fractional and integral effects.
\vskip 1cm


\vfill
CERN-TH 7344

June 94

cond-mat/9406119

\eject

\countdef\pageno=0 \pageno=1
\newtoks\footline \footline={\hss\tenrm\folio\hss}
\def\folio{\ifnum\pageno<0 \romannumeral-\pageno \else\number\pageno \fi}
\def\advancepageno{\ifnum\pageno<0 \global\advance\pageno by -1
\else\global\advance\pageno by 1 \fi}

\noindent There has been a lot of interest in the last years on the Quantum
Hall Effect (QHE)\ref{1}. The integral effect (IQHE) has been given simple
explanation\ref{2}, but the fractional effect (FQHE) is more involved and
requires the concept of collective behavior of the electrons\ref{3}. In
particular, Jain\ref{4} used a unified scheme in order to explain both effects,
 since the explanation of the integral effect was believed to be described in
terms of non-interacting electrons\ref{1}, while the fractional effect
includes the above mentioned collective phenomenon. The idea is to introduce
copies of the electron, and consider condensations of such fields\ref{4}.

Further understanding  can be achieved once one verifies that chirality plays
an essential role in the explanation of the Hall effect, as stressed
recently\ref{5,6,7,8}. Our purpose here is to provide a field theoretical
approach to the quantum Hall effect, in such a way that the collective
electron phenomenon is naturally accomodated, being just a consequence of the
description of electrons squeezed in a small space region, and the chiral
nature of the interaction\ref{5-8}, leading to chiral anomaly, which together
with the soliton behavior inherited from the collective effect, naturally
accomodates the fractional effect, and gives, as a byproduct, the integral
effect as a particular case. We refrain from reviewing the subject, refering
to the vast literature\ref{1,9}. However, we have to explicitly quote some
results which will be used in the following, as well as refer to as a check of
the correctedness of the procedure.

We start out of the Lagrangean density of four-dimensional QED, which reads
$$
{\cal L}= \overline \Psi_i \Gamma_\mu i \partial _\mu \Psi_i + e \overline \Psi
_i \Gamma^\mu A_\mu \Psi_i \quad, \quad i=1,\dots , N\quad ,\eqno(1)
$$
where the gamma matrices $\Gamma^\mu$ and the $\Psi$ field are four-dimensional
 and read
$$
\Gamma^0 = \left( \matrix{0&1\cr -1& 0\cr}\right)\quad ,
\quad \Gamma^a = \left( \matrix{0&\sigma^a\cr -\sigma^a& 0\cr}\right)\quad ,
\quad \Psi = \left( \matrix{\chi\cr \eta\cr}\right)\quad ,\eqno(2)
$$
$\sigma^a $ are the usual Pauli matrices with $a=1,2,3$ and $\chi$ and $\eta$
are bispinors. In this notation the Lagrangian is
$$
\eqalignno{
{\cal L} &= \chi^+\partial_0\chi +\eta^+\partial_0 \eta -
\eta^+\sigma^a\partial_a\eta -  \chi^+ \sigma^a \partial^a \eta + \chi^+\chi
A_0 + \eta^+\eta A_0 - e(\chi^+ \sigma^a\chi + \eta^+\sigma^a\eta)A_a\cr
& = \overline\chi\not \! \partial\chi + \overline \eta\not \!\partial \eta
+ (\overline \chi \gamma^\mu\chi + \overline \eta \gamma^\mu\eta)A_\mu +
\chi^+\sigma_{_M}\partial_{_M}\chi + \eta^+\sigma_{_M}\partial_{_M}\chi\cr
& = \overline \psi _{_I}  \not \!\partial \psi_{_I} + \overline \psi_{_I}
\gamma^\mu \psi_{_I} A_\mu + {\rm M\!-\!derivatives}\quad , \quad I= 1,2\,\,
{\rm and}\,\,  M=2,3\quad .&(3)\cr}
$$
In the above we have used the following notation: The spinor field
$\psi_{_{I=1}}$ stands for $\chi$, the first bispinor, while $\psi_{_{I=2}}$
corresponds to $\gamma_1\eta (t,-\vec x)$, and the two-dimensional gamma
matrices are $\gamma_0=\sigma^3$,   $\gamma_1=\sigma^2$, and $\gamma_5=
\gamma_0\gamma_1$; the slash derivative is $\not\!\partial = \gamma^\mu\partial
_\mu$. Therefore, for $\Gamma_\mu$ the index $\mu$ goes from 0 to 3, while for
$\gamma_\mu$ it takes the values 0,1. We neglect the derivatives in the 2 and 3
 directions ($M$),considering only the two-dimensional problem.
\eject

Integrating over the gauge field $A_\mu (\mu = 0, \cdots , 3)$ we get
$$
{\cal L}=\overline\psi_j i\not\!\partial\psi_j+ \sum_{\mu,\nu=0}^{1}\overline
\psi_j\gamma^\mu\psi_j D_{_{\mu\nu}}\overline\psi_k\gamma^\nu\psi_k+\overline
\psi_j\psi_j D_{_{33}}\overline\psi_k\psi_k +\overline \psi_j\gamma^5\psi_j
D_{_{22}}\overline\psi_k\gamma^5\psi_k\quad ,\eqno(4)
$$
where $j,k =1,\dots , 2N$ and $D_{_{\mu\nu}}$ (for all the indices $\mu, \nu =
0, \cdots , 3$), is the gauge field propagator  and  may be approximated by an
effective constant in  the strong coupling regime. We are left with the
affective Lagrangean
$$
{\cal L} = \overline \psi_ j i\not \! \partial \psi_j + g_{eff} \overline\psi_j
\gamma^\mu \psi_j\overline\psi_k\gamma_\mu\psi_k  + g_{eff}\left[ (\overline
\psi_j\psi_j)^2 - (\overline\psi_j\gamma_5 \psi_j)^2 \right]\quad ,\eqno(5)
$$
where the effective current vertex corresponding to $\mu =2$ in two dimensions
is $\overline\psi\gamma_5 \psi$ and for $\mu=3$, it is  $\overline\psi\psi$.
We find also a trivial Thirring interaction, which is neglected in the
following.  The  electromagnetic field is responsible for the internal
interactions in the cristal, and for the formation of the two-dimensional
layer. Now the effective  two-dimensional interaction is the chiral Gross-Neveu
(CGN) theory, which has been much  studied in the literature (see [10] and
references therein). Since the electromagnetic field has been used in order to
form the system, we shall for the time being disregard its interaction. Later
we have to reconsider it, in order to know the interaction of the sample with
the external fields responsable for the Hall effect. As far as the CGN
interaction is concerned, we recall the following facts. The solution of such a
 theory is given in terms  of solitons $\hat\psi_j$, which satisfy the
non-linear relation\ref{10,11}
$$
{\widehat\psi_j}^+ \sim \epsilon_{jj_1\cdots j_{2N-1}}\widehat\psi_{j_1}\cdots
\widehat\psi_{j_{2N-1}}\quad ,\eqno(6)
$$
where on the right hand side a suitable redefinition of the Klein\ref{10}
factor and the normal product prescription are required.

Such solitons are chargeless fields, since the $U(1)$-charge has been separated
in order that the above relation be valid. In fact, the gauge interaction
has been used up in the quantum gauge degrees of freedom, in order that the
two-dimensional infra-red behavior does not cause problems. As shown by
Wen\ref{12}, the excitations responsable for the QHE contain several branches.
In fact, the excitations in each branch carry a fractional charge. Such edge
excitations form the Fermi liquid. However, as stressed by Wen, the fermions in
the Fermi liquid are not the electrons, but describe the edge excitation, a
possibility opened by the boson/fermion duality in two dimensions. This means
that the Fermi liquid is built of solitons,  described by generalized
statistics\ref{13,14}.

The next procedure to be followed concerns the interaction of the solitons with
the external gauge fields. The gauge interaction of the degenerate solitons
displaying an apparent $SU(2N-1)$ symmetry, with the gauge field is
fundamentally important in order to explain the effect. The discussion of gauge
 invariance in this context is given in [15].  From the fact that we have a
bound state structure, or collective approach of the type  pictured by eq. (6),
 we can describe the physical electron field by the l.h.s. of (6) times the
$U(1)$ factor ($\psi_j = {\rm e}^{i \chi }\widehat\psi_j $), since it is such
field which has physical interpretation. This means that we can distribute the
charge equally among the soliton fields, and try to find an effective
Lagrangean which displays the desired features of the Hall effect. As a pure
two-dimensional theory, the chiral Gross Neveu model presents fields with
generalized statistics. As a four-dimensional theory we interprete its meaning
as the fact that only bound states of the form (6) describe the physical
electron. Therefore, we suppose that the physical electron is of the above
form, multiplied by the convenient exponential representing the $U(1)$ charge,
that is
$$
\left(\hat \psi {\rm e}^{-i\chi/n}\right)^+ = {\rm e}^{i\chi}\,
\epsilon_{jj_1\cdots j_n}\left(\hat \psi_{j_1} {\rm e}^{-i\chi/n}\right)\cdots
\left(\hat \psi_{j_n} {\rm e}^{-i\chi/n}\right)\quad ,\eqno(7)
$$
and each of the solitons carries a charge ${e\over n}$.  Such a charge can be
traced back to the incompressibility, as discussed by Laughlin, due to the
Schieffer counting argument\ref{16}.

We have started with $N$ fields. In the reduction scheme, due to the fact that
four-dimensional electrons have four components, while two-dimensional
electrons
have two, there is a doubling in the number of components. Therefore the
number of independent fields  is $n=2N-1$. This mirrors the relativistic
content of the model, that is, although the spin does not play a direct role
in the QHE, it enters through the above doubling of components. However, the
incompressible Fermi fluid picture of Laughlin, has a consequence that the
excitations are anyons with fractional statistics and charge\ref{13,14}.

We now come to the effective action describing the solitonic fields. It may be
obtained by the minimal coupling of such fields with the external
electromagnetic field as
$$
{\cal L}_{eff}= \overline \psi_j i \not \!\partial \psi_j + \hat e\overline
\psi_j\not \!\! A P_-\psi_j + {\hat a \hat e^2\over 4\pi}A_\mu A^\mu + \left(
\psi \! - \! {\rm self \; interactions} \right)\quad ,\eqno(8)
$$
where we followed [5,6] and introduced the interaction of the electromagnetic
field with the current of definite chirality, that is $\overline \psi_j\gamma
^\mu P_-\psi_j\equiv\overline\psi_j\gamma^\mu{1-\gamma_5\over 2}\psi_j$.
Above, the  charge of the soliton is $\hat e\!=\!{e\over 2N-1}$, and $\hat a \!
=\!(2N\!-\!1)a$ represents the regularization ambiguity\ref{17} (see below).
In the QHE  chirality plays a fundamental role. It is well known that chiral
gauge interactions are anomalous\ref{10,17,18}, and two-dimensional chiral QED
with anomalous breakdown of gauge invariance\ref{17} admits exact solutions in
a positive metric Hilbert space respecting unitarity, provided the  parameter
$a$, defined above (see [17]) is restricted to $a\ge 1$. In fact, the JR\ref
{17} term $ae^2A_\mu^2$ takes account of the ambiguity in the regularization
procedure resulting from the lack of gauge invariance. It may also be obtained
as the massless limit of the Proca theory\ref{10}.

This is our proposal for the effective field theory Lagrangean to describe the
QHE. Therefore we arrive at the chiral $QED_2$ interaction, together with
chiral Gross-Neveu type self-interaction for the Fermi fields, as well as
the constraint (6). Since the essence of the chiral Gross-Neveu
self-interaction is to imply the relation (6) characteristic of the
two-dimensional theory, while the Hall effect itself depends on the relation
between the current and the external gauge field, we simplify matters,
considering chiral $QED_2$ as the model for the QHE, together with relation
(6). (In fact, in the presence of impurities the free electron picture holds
true\ref{13,14}).

An effective bosonic theory is obtained by means of the computation of the
fermionic determinant in two dimensions, through the Polyakov-Wiegman
formula\ref{19}. In such a case, integration over the fermions leads to an
external gauge field dependent partition function of the form\ref{10}
$$
{\rm e} ^{iW^L[A]} = {\rm e}^{i \int {\rm d}^2 x\, {\hat a \hat e^2 \over 4\pi}
A_\mu^2}\int {\cal D}h \, {\rm e} ^{i\gamma_1 [A,h^{-1}]}\quad ,\eqno(9)
$$
where $h$ is a bosonic field, $L$ means that we are considering the interaction
of the left moving currents with the gauge field as in (8) and $\gamma_1
[A,h^{-1}]$ is the effective bosonic action satisfying a 1-cocycle
condition\ref{10}. It is defined in terms of the WZW functional\ref{20} as
$$
\eqalignno{
\gamma_1 [A,h] &= \Gamma[h V^{^{-1}}] - \Gamma[V^{^{-1}}]\quad , &(10a) \cr
&= \Gamma [h] - {i\hat e\over 4\pi} \int {\rm d}^2 x \, A^\mu h^{^{-1}}
(\partial_\mu -\widetilde \partial_\mu )h \quad ,&(10b)\cr
\noalign {\hbox {where the WZW functional is given by the expression }}\cr
\Gamma[h] & = {1\over 8\pi}\int {\rm d}^2 x \, \tr \partial _\mu h^{^{-1}}
\partial
^\mu h - {i\over 4\pi}\tr\epsilon^{\mu\nu}\!\int_0^1\!{\rm d} r\int{\rm d}^2
x\,
\widetilde h^{^{-1}}\partial _r \widetilde h \widetilde h ^{^{-1}} \partial_\mu
\widetilde h \widetilde h^{^{-1}} \partial_\nu \widetilde h\, . &(10c)\cr}
$$
In the above we chose $A_-={i\over\hat e}V^{^{-1}}\partial_-V$ and $A_+=0$. In
the WZW functional $\Gamma[h]$ the general $\widetilde h(r,x)$ field is an
extension of the field $h(x)=\widetilde h(1,x)$ to a manifold containing the
Minkowski space as a bondary with $\widetilde h(0,x)=1$. In the abelian case $(
h_j={\rm e}^{2i\phi_j})$ the effective action boils down to\newfoot{\ref{
\dagger}} {As a matter of fact, we are describing a torus $U(1)^{2N-1}$.}
$$
\gamma_1 [A,\phi_j] = \int {\rm d}^2 x \, {1\over 2}\sum_j (\partial_\mu
\phi_j)^2 + {\hat e \over 2\pi} \int {\rm d}^2 x \, \sum_j A^\mu (\partial
_\mu -\widetilde \partial_\mu )\phi_j \quad .\eqno(11)
$$
Thus the bosonized form of two-dimensional chiral QED with flavour reads
$$
{\cal L} = {1\over 2} \sum_j(\partial_\mu \phi_j)^2 - {1\over 4}F_{\mu\nu}^2
+ {\hat e\over 2\pi}A^\mu \sum_j(\partial _\mu - \widetilde \partial_\mu)\phi_j
+ {\hat a\hat e^2\over 4\pi}A_\mu A^\mu \quad .\eqno(12)
$$
We could also obtain this action with the adiabatic principle of form
invariance\ref{10}. The equations of motion obtained from (12) are
$$
\eqalign{
\partial_\mu F^{\mu\nu} + {1\over 2\pi}\sum_i(\partial^\nu - \widetilde
\partial^\nu)\phi_i + {\hat a\hat e\over 2\pi}A^\nu &=0\quad ,\cr
\dal \phi_i + {\hat e\over 2\pi}(\partial_\mu - \widetilde
\partial_\mu)A^\mu &=0\quad .\cr }\eqno(13)
$$

Consider now the (classically conserved) currents
$$
\eqalign{
J_L^\mu & = {\hat e\over 2\pi}\sum_j (\partial^\mu - \widetilde \partial ^\mu)
\phi_j + {\hat a\hat e^2\over 2\pi}A^\mu \quad ,\cr
J_{i R}^\mu & = - (g^{\mu\nu} + \epsilon^{\mu\nu})(\partial_\nu \phi_i
-\hat e A_\nu)\quad .\cr}
\eqno(14)$$
In the quantum theory we have conservation of the right-moving current $J^\mu
_{iR}$, as expected,
$$
\partial_\mu J_{i R}^\mu =0 \quad , \eqno(15)
$$
but the left current is anomalous! Indeed, we have
$$
\partial _\mu J_L^\mu = (2N-1){\hat e^2\over 2\pi}\left[ (a-1)\partial_\mu +
\widetilde \partial_\mu\right] A^\mu \quad .\eqno(16)
$$

The fact that the  ``right" current is conserved at the quantum  level, while
the ``left" current is anomalous is a very ``efficient" description of the
physics of the QHE, and may also be related to its description in terms of
chiral bosons interacting with matter fields\ref{5,6}. In such a case, we also
have higher algebras describing the theory\ref{21}.

The question whether the anomalous conservation is compatible or not with the
equation of motion has been answered in a series of papers (see [10]) for an
extensive list). In the case we are considering a functional integration over
the gauge field $A_\mu$, i.e., if the electromagnetic field is fully quantized,
the r.h.s. of (16) vanishes. Here we are considering an external gauge field -
the one originating the Hall effect - therefore the r.h.s. of (16) does not
vanish, but there is no contradiction with the equations of motion. For the
applications on the QHE we have an $(2N-1)$-plet of fields, therefore
$$
\eqalign{
\partial^\mu J_\mu^L & = (2N-1){e^2\over 2\pi(2N-1)^2}\left[(a-1) \partial_\mu
+ \widetilde \partial_\mu\right] A^\mu\quad ,\cr
& = {e^2\over 2\pi(2N-1)}\left[(a-1) \partial_\mu
+ \widetilde \partial_\mu\right] A^\mu\quad .\cr}\eqno(17)
$$

Since $\widetilde \partial_\mu A^\mu = \epsilon_{\mu\nu} \partial^\nu A^\mu =
-{1\over 2}\epsilon_{\mu\nu}F^{\mu\nu}$, we read the Hall conductivity from the
anomaly coefficient. The first term, namely $(a-1)\partial_\mu A^\mu$ is a
pure two-dimensional effect, and does not appear if the external gauge field
describes the $\vec E\, ,\,\vec B$ usual system. Therefore, the (minimal) Hall
 conductivity is given by the coefficient of the anomaly\ref{22},
$$
\sigma_H = {e^2\over 2\pi(2N-1)}\quad .\eqno(18)
$$

We conclude remarking once more that the current explanation of the quantum
Hall
conductance is naturally accomodated in terms of the two-dimensional field
theoretic language. The fact that we have a two-dimensional relativistic
field theory instead of a three-dimensional (non-relativistic) theory has been
discussed elsewhere\ref{23}. The description of this phenomenon in terms of
a conformal field theory has also been developed in references [24, 25, 26].

We should also quote that copies of a $U(1)$ Kac-Moody algebra are obtained
by the currents $J_\mu^R$. Indeed,
$$
\left[ J_{i-}^R(t,x), J_{j-}^R(t,y)\right]={1\over 2\pi} \hat e^2
\delta_{ij}\delta^\prime (x-y)\quad ,\eqno(19)
$$
which should be compared to [12]. Moreover, the $W_\infty$ algebra obtained
for the non-relativistic electron gas description\ref{23} can be understood
as the algebra of higher spin chiral current studied is [27] for the abelian
case, and in [28] in the non-abelian case. Such higher-dimensional algebras
underline the models considered\ref{21}.

Although we did not discuss the issue of the hierarchy of the different filling
factors, it should be clear at this point that once we have gotten (18), and
consequently the simplest filling factor $\nu =1/3$ (for $N=2$), the arguments
used in [29] now apply.
\vskip .5cm
\noindent
This work was partially supported by CAPES (E.A.), Brazil, under contract
No. 1526/93-4 and by CNPq (M.C.B.A.), Brazil, under contract No. 204220/77-7.
\vskip 1cm
\noindent  {\bf References}
\vskip .5cm
\refer [[1]/R.E. Prange and S.M. Girvin, ed., {\it The Quantum Hall Effect},
Springer, Berlin, 1990 ]

\refer [[2]/R.B. Laughlin, Phys. Rev. {\bf B23} (1981) 5632-5633]

\refer[[3]/R.B. Laughlin, Phys. Rev. Lett. {\bf 50} (1983) 1395-1398]

\refer[[4]/J.K. Jain, Phys. Rev. Lett. {\bf 63} (1989) 199; Phys. Rev.
{\bf B40} (1989) 8079]

\refer[[5]/L. Chandar, Fractional Quantum Hall Effect from Anomalies in WZNW
Model, Preprint SU-4240-569, february, 1994]

\refer[[6]/A.P. Balachandran, L. Chandar and B. Sathhiapalan, Duality and the
Fractional Quantum Hall Effect, Preprint SU-4240-578, May, 1994]

\refer[[7]/M. Stone, Phys. Rev. {\bf B42} (1990) 8399-8404]

\refer[[8]/X-G. Wen, Phys. Rev. {\bf B41} (1990) 12838-12844]

\refer[[9]/M. Stone, Ed., {\it Quantum Hall Effect}, World Scientific, 1992]

\refer[[10]/E. Abdalla, M.C.B. Abdalla and K. Rothe, {\it Non-perturbative
Methods in Two-\-dimen\-sional Quantum Field Theory}, World Scientific, 1991]

\refer[[11]/R. K\"oberle, V. Kurak, J.A. Swieca, Phys. Rev. {\bf D20} (1979)
897; E {\bf D20} (1979) 2638]

\refer[[12]/X-G. Wen, Phys. Rev. Lett. {\bf 64} (1990) 2206; NSF-ITP-89-157
preprint, (unpublished)]

\refer[[13]/F. Wilczek, Ed., {\it Fractional Statistics and Anyon
Supercondutivity}, World Scientific, 1990]

\refer[/S. Forte, Rev. Mod.  Phys. {\bf 64} (1992) 193]

\refer[[14]/R.B. Laughlin, Fractional Statistics and QHE, in {\it Fractional
Statistics and Anyon Supercondutivity}, Ed., F. Wilczek, World Scientific,
1990]

\refer[[15]/R. Tao, Y.S. Wu, Phys. Rev. B30 (1984)1097.]

\refer[[16]/R.B. Laughlin Phys. Rev. Lett. 50 (1983)1395; W.P. Su and
J.R. Schrieffer Phys. Rev. Lett. 46 (1981) 738]

\refer[[17]/R. Jackiw and R. Rajaraman, Phys. Rev. Lett. {\bf 54} (1985) 1219]

\refer[[18]/K. Harada and I. Tsutsui, Phys. Lett. {\bf B183} (1987) 311]

\refer[/K. Harada, Nucl. Phys. {\bf B329} (1990) 723]

\refer[/H.O. Girotti and K. Rothe, J. Mod. Phys. {\bf A4} (1989) 3041]

\refer[[19]/A.M. Polyakov and P.B. Wiegman, Phys. Lett. {\bf 131B} (1983) 121;
{\bf 141B} (1984) 223]

\refer[[20]/J. Wess and B. Zumino. Phys. Lett. {\bf 37B} (1971) 95;]

\refer[/E. Witten, Commun. Math. Phys. {\bf 92} (1984) 455]

\refer[[21]/E. Abdalla, M.C.B. Abdalla, L. Saltini and A. Zadra, work in
progress]

\refer[[22]/C.G. Callan and J.A. Harvey, Nucl. Phys. {\bf B250} (1985) 427-436]

\refer[[23]/A. Capelli, C.A. Trugenberger and G.R. Zemba, Nucl. Phys. {\bf
B396} (1993) 465-490]

\refer[[24]/S. Fubini, Mod. Phys. Lett. {\bf A6} (1991) 347]

\refer[[25]/M. Stone, Int. J. Mod. Phys. {\bf B5} (1991) 509]

\refer[[26]/S. Fubini and C.A. L\"utken, Mod. Phys. Lett. {\bf A6} (1991) 487;]

\refer[/G.V. Dunne, A. Lerda and C.A. Trugenberger, Mod. Phys. Lett. {\bf A6}
(1991) 2819;]

\refer[/C. Cristofano, G. Maiella, R. Musto and F. Nicodemi, Phys. Lett. {\bf
B262} (1991) 88; Mod. Phys. Lett. {\bf A6} (1991) 1779, 2985]

\refer[[27]/G. Sotkov and M. Stanishkov, Off-critical $W_\infty$ and Virasoro
algebras as dynamical symmetries of integrable models, NATO Workshop,
``Integrable Quantum Field Theories", Como, Italy, 1992]

\refer[[28]/E. Abdalla, M.C.B. Abdalla, G. Sotkov, M. Stanishkov, Off-critical
current algebras, Int. J. Mod. Phys. A, to appear]

\refer[[29]/B. Blok and X.G. Wen, Phys. Rev. B10 (1991) 8337;  F.D.M. Haldane,
 Phys. Rev. Lett. 51 (1983) 605]
\end